%% file: sample-sigconf.tex
\documentclass[sigconf]{acmart}
\renewcommand\footnotetextcopyrightpermission[1]{} 
\usepackage{fancyhdr}
\usepackage{booktabs} 

\setcopyright{rightsretained}

\acmDOI{}

\acmISBN{}

\acmConference[]{Impact of Feature Selection on {Micro-Text} Classification}
\acmYear{2017}

\begin{document}
\title{Impact of Feature Selection on {Micro-Text} Classification}

\author{\textbf{Ankit Vadehra}}
\affiliation{\institution{University of Waterloo}}
\email{avadehra@uwaterloo.ca}

\author{\textbf{Maura R. Grossman}}
\affiliation{\institution{University of Waterloo}}
\email{maura.grossman@uwaterloo.ca}

\author{\textbf{Gordon V. Cormack}}
\affiliation{\institution{University of Waterloo}}
\email{gvcormack@uwaterloo.ca}

\renewcommand{\shortauthors}{Vadehra, Grossman \& Cormack}
%
%


\begin{abstract}
Social media datasets -- especially Twitter tweets -- are popular in the field of text classification. Tweets are a valuable source of micro-text (sometimes referred to as ``micro-blogs''), and have been studied in domains such as sentiment analysis, recommendation systems, spam detection, clustering, among others \cite{atc}. Tweets often include keywords referred to as ``Hashtags'' that can be used as labels for the tweet. Using tweets encompassing 50 labels, we studied the impact of word versus character-level feature selection and extraction on different learners to solve a multi-class classification task. We show that feature extraction of simple character-level groups performs better than simple word groups and  pre-processing methods like normalizing using Porter's Stemming and Part-of-Speech (``POS'')-Lemmatization.
\end{abstract}


\maketitle

\input{samplebody-conf}

\bibliographystyle{ACM-Reference-Format}
\bibliography{sigproc}
\end{document}

%% file: samplebody-conf.tex
\section{Introduction}

There has been an explosion in the use of social-networking platforms like Twitter, Facebook, Instagram, and blogs, to name but a few. Twitter restricts posts to 140 characters each, which makes them an excellent source for micro-text. Tweets often contain Hashtags, denoted by keywords preceded by the \# symbol. For example, a user discussing the 2016 U.S. presidential election might write ``\#elections16'' in his tweet. This allows the tweet to be added to a virtual set of all tweets talking about the same topic. A simple search for ``\#elections16'' allows a user to monitor the ongoing public debate on this topic; it also acts as a label for the tweet and can result in the Hashtag being added to the trending topics if many people are talking about the issue. 

Automatic Hashtag suggestions can be provided to users writing new tweets using a classifier trained on trending Hashtags. This can help the user add his views to the ongoing discussion of a topic that is of widespread public interest. Therefore, this is a highly desirable feature that could be used in a Hashtag recommendation system.

In this work, we compare the performance of various learners with minimal tuning on a fairly large dataset of tweets to determine the impact of different features on prediction accuracy.  Previous work in this area has looked at comparing different classification techniques for text classification of tweets \cite{tclass}, and tweet classification on the basis of different heuristics such as user profile, post time, profile time, and textual content \cite{dclass}. Prior work using the word-level bigram features for baseline classifiers showed them to be more accurate than individual words \cite{bigram}. Others have tried using multiple word-normalization techniques for better generalization and gains in accuracy prediction, but they have not shown any gains in performance \cite{wnorm}.

In the following sections, we present further detail. Section 2 describes the different feature selection and extraction techniques.  Section 3 addresses the different learners ({\em i.e.}, machine-learning models) used for classification, along with their parameters. Section 4 explains the experimental set-up.  We then turn to the results of our experiment in Section 5, and present our conclusions and suggestions for future work to expand on this study in Sections 6 and 7, respectively.

\section{Feature Selection}
Feature selection is one of the most crucial aspects in designing a classifier. We select N-Gram groups of varying length at the character and word level as features. The frequency count (``FC'') and TF-IDF score for the selected features is extracted and used to train the classifier model.

We vary the N-Gram group range until the prediction accuracy ceases to increase or the system runs out of memory. Separate runs for N-Gram groups at the character/word level using different classifiers is performed and the results are shown for evaluation. We use individual and combined features, which is further explained in the Results section below.
\section{Classifier Models}
For testing the impact of different features, we trained basic machine-learning models that are considered baseline for different textual classification tasks \cite{bigram}. We consider the following models with fixed parameters:\\
$\bullet$ \textbf{Support Vector Machine (``SVM''):} SVM's have been used in various text classification tasks and have been shown to give good results \cite{SVMexp}. SVM tries to optimize the equation: $1/2||w||^2 + C \Sigma x_i$, where C is the slack variable, which allows a trade-off between the hyperplane margin and the number of misclassifications. We use an SVM with a linear kernel, C=1, a squared-hinge loss function, and L2 loss penalty \cite{sklearn}.

$\bullet$ \textbf{Multinomial Naive Bayes (``MNB''):} Naive Bayes is the most common classifier used in text classification. Relying on a simple probabilistic model, it learns easily and is used widely in practice. It is based on the Bayes rule: $P(c_i|i) = \frac{P(i|c_i)P(c_i)}{p(i)}$, where $P(c_i|i)$ = probability that item $i$ belongs in class, $c_i$. $P(c_i)$ = (\# of training items for class $c_i$ / total number of training examples). $P(i)$ = normalizing factor, which can be omitted since it is the same for all classes. $P(i|c_i)$ = probability estimate that the item $i$ was present in the training of $c_i$. We use Multinomial variant with Laplace smoothing parameter = 1\cite{sklearn}.
\section{Experimental Setup}
\subsection{Dataset}
The dataset was gathered using the basic version of the Twitter sample streaming API, which returns a small random sample of all public posts. Tweets for the month of September, 2016 were used. This set was reduced by selecting only those tweets where the language was English and there was at least one Hashtag present. We removed all recurring tweets. This resulted in a set of $1,602,604$ unique Hashtags and $19,611,453$ tweets. Out of these, we chose the top $50$ Hashtags as class labels and the corresponding set of tweets containing those Hashtags. This resulted in a set of $964,889$ tweets, spread across the $50$ selected Hashtags. The distribution of the Hashtag labels, along with the number of tweets is supplied in Table \ref{table:freq}.
\begin{table}
  \caption{Frequency of Selected Hashtags}
  \label{table:freq}
  \begin{tabular}{p{3.2cm}p{2.4cm}p{2.4cm}}
    \toprule
    Hashtag :Tweet Count&&\\
    \midrule
nowplaying:76618&debatenight:66914&android:60838\\
job:52683&news:41459&travel:39647\\
hiring:38776&trump:32076&giveaway:30659\\
marketing:26338&music:25789&neverforget:23982\\
fashion:22590&photography:18818&love:18330\\
iran:18286&emmys:16517&hillary:15360\\
tech:15283&saudi:14947&socialmedia:14295\\
blacklivesmatter:13514&instagram:13021&uber:12951\\
health:12752&onedirection:12735&tbt:12530\\
nyc:12306&photo:12293&competition:12241\\
lyft:11837&youtube:11634&pokemongo:10511\\
iphone:10361&ios10:9753&football:9517\\
startup:9399&syria:9399&sale:9199\\
leadership:8949&entrepreneur:8799&books:8221\\
mondaymotivation:8058&ipad:8042&uk:8031\\
code:7854&fitness:7788&gaming:7723\\
us:7673&apple:7593&\\
  \bottomrule
\end{tabular}
\end{table}
\subsection{Data Cleaning}
The dataset was minimally cleaned to train the classifiers. The Hashtag was removed from each tweet and was used as the class label. For example, the tweet ``\textit{want to work at robert half technology? we're in nc click for details. \#hiring}'' is converted to ``\textit{want to work at robert half technology were in nc click for details}.'' The Hashtag ``\#hiring'' is used as the class label for the tweet. Apart from this, the following additional cleansing steps were taken:\\
$\bullet$ Convert all unicode symbols into ascii. Some accents and brackets were removed.\\
$\bullet$ Remove all whitespace characters, apart from spaces, so that newline, tabs etc. are removed.\\
$\bullet$ Remove the keyword ``RT'' from the Tweets. ``RT'' denotes that the tweet is a retweet.\\
$\bullet$ Convert the whole tweet into lowercase.\\
$\bullet$ Remove all URLs/hyperlinks from the tweets.\\
$\bullet$ Remove all characters other than alphanumerics and spaces.\\
$\bullet$ Convert occurrence of multiple spaces into a single space.\\
$\bullet$ Replace the occurrence of two or more characters in succession by the character itself. For example, ``\textit{i'm so happyyyyyyyy...}'' is converted to ``\textit{im so hapy}.''\\
$\bullet$ Remove starting and ending trailing whitespace.\\
Using these steps, a relatively clean dataset was generated.
\subsection{Research Question}
Using this dataset, we now wanted to determine the impact of different character/word-level features using different feature-extracted scores on prediction accuracy. The main question we wished to answer was:

\textit{Are the character-level features comparable to or better than the word-level and pre-processed (Stemmed/Lemmatized) features for accuracy?}

In various text classification tasks, features are typically selected at the word level, which begs the question \textit{why not use character-level features?} One trade-off is that when we move away from word-level features, we may lose the contextual value of the terms. Hence, we compared word-Lemmatization(using Parts-of-Speech tag) and Stemming results with character-level results to answer this question. As an extension, we also compared the results using two different feature-extraction methods, namely: Frequency Count (``FC'') and TF-IDF, to determine the impact on the different learners.

\subsection{Baseline}
For all classifiers, the baseline result was taken as the prediction accuracy after training on individual words using Frequency Count as a feature extraction step, since this is a common classification technique.

\subsection{Evaluation Metrics}
The evaluation was performed using total accuracy.\\
Total Accuracy = $\frac{\# \ of \ correct \ predictions}{\# \ of \ correct \ + \ incorrect \ predictions}$\\
A tweet may have more than one Hashtag, and a prediction was deemed to be correct if it identified any one of the Hashtags in that tweet.

\subsection{Test and Training Set}
The total dataset consisting of 964,889 Tweets was randomly sorted and then split using a 70:30 ratio, where 70\% = Training Set and 30\% = Testing set.

\section{Results}
\begin{figure}
\includegraphics[width=2.756in,height=2.12in,scale=0.28]{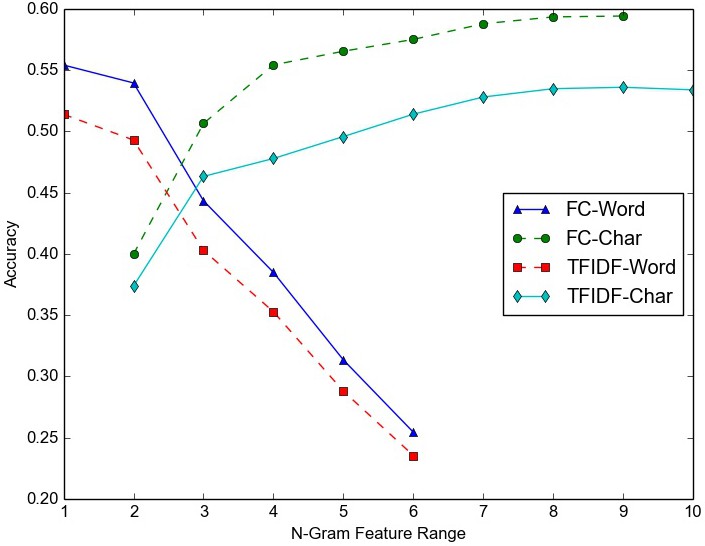}
\caption{MNB with Individual N-Gram.}
\label{fig:1}
\end{figure}

\begin{figure}
\includegraphics[width=2.756in,height=2.12in,scale=0.28]{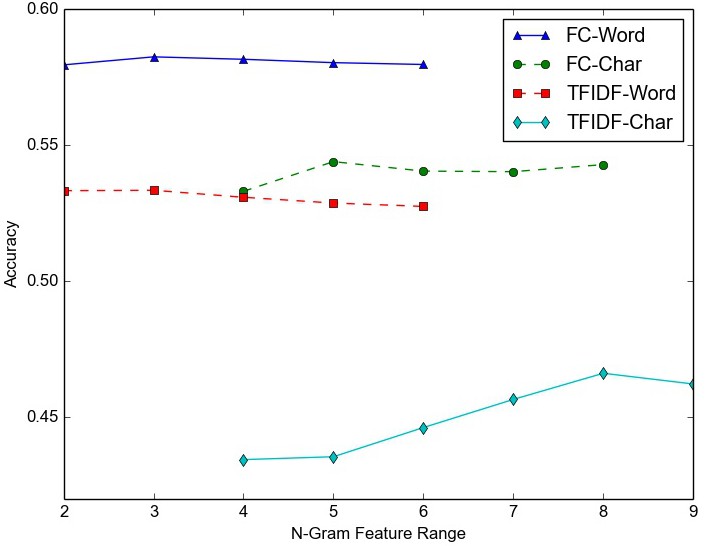}
\caption{MNB with Combined N-Gram.}
\label{fig:2}
\end{figure}

\begin{figure}
\includegraphics[width=2.756in,height=2.12in,scale=0.28]{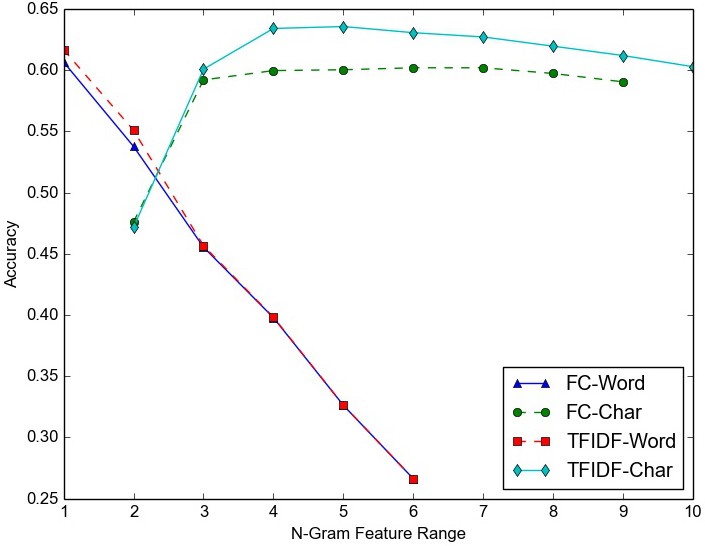}
\caption{SVM with Individual N-Gram.}
\label{fig:3}
\end{figure}

\begin{figure}
\includegraphics[width=2.756in,height=2.12in,scale=0.28]{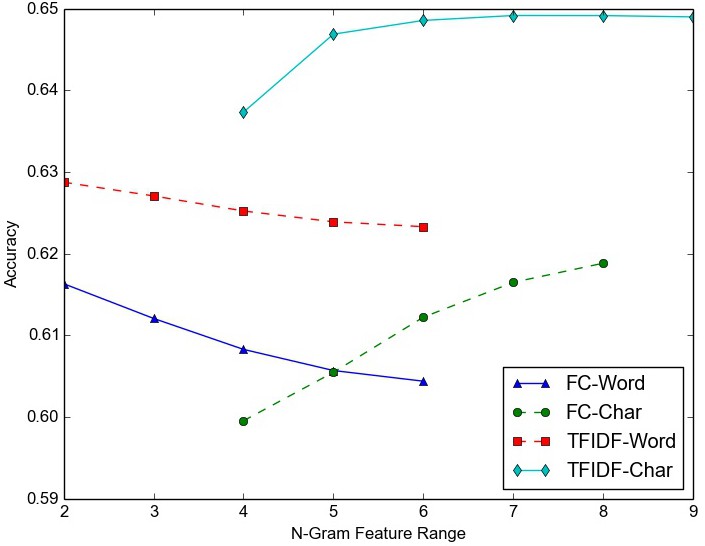}
\caption{SVM with Combined N-Gram.}
\label{fig:4}
\end{figure}

\begin{figure}
\includegraphics[width=2.756in,height=2.12in,scale=0.28]{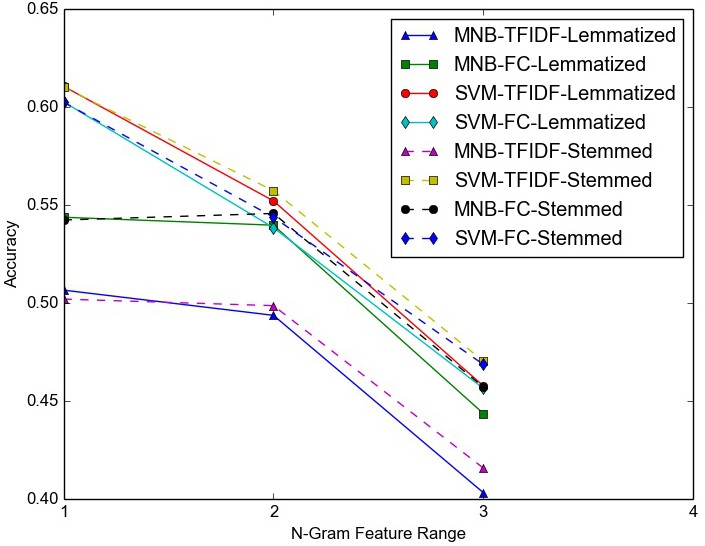}
\caption{Post-Processing with Individual N-Gram.}
\label{fig:5}
\end{figure}

\begin{figure}
\includegraphics[width=2.756in,height=2.12in,scale=0.28]{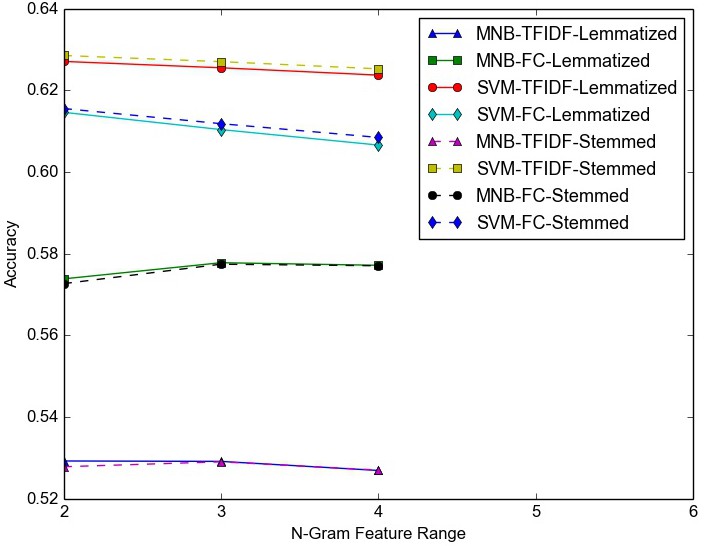}
\caption{Post-Processing with Combined N-Gram.}
\label{fig:6}
\end{figure}

Multiple runs were performed for each of the selected feature/model/extraction methods. We denote by $(i,j)$ the use of all N-grams (either word or character) from $N=i$ through $N=j$.  We report the following combinations:\\
$\bullet$ \textbf{Individual N-Gram Group: } N represents the N-Gram group $(N,N)$ taken individually. For example 1: unigram, 2: Bigram, 3: trigram, and so on.\\
$\bullet$ \textbf{Combined N-Gram Group: } N represents the combinations $(1,N)$. For example 1: Unigram, 2: Unigram + Bigram, 3:Unigram + Bigram + Trigram, and so on.\\
$\bullet$ \textbf{Learner Models: }SVM = Support Vector Machine, MNB = Multinomial Naive Bayes.\\
The highest-accuracy group for each learning and feature method is tabulated in Table \ref{table:final}.
\begin{table}
  \caption{Highest Accuracy Groups for Different Features}
  \label{table:final}
  \begin{tabular}{lll}
    \toprule
    Model and Features&N-Gram Group&Accuracy\\
    \midrule
MNB-FC&word(1,3)&0.582\\
SVM-FC&word(1,2)&0.616\\
LR-FC&char(1,4)&0.624\\
MNB-TFIDF&char(8,8)&0.534\\
SVM-TFIDF&char(1,7)&0.649\\
LR-TFIDF&char(1,4)&0.614\\
MNB-Stemmed&FC-(1,4)&0.577\\
SVM-Stemmed&TFIDF-(1,2)&0.628\\
MNB-Lemmatized&FC-(1,3)&0.578\\
SVM-Lemmatized&TFIDF-(1,2)&0.627\\
  \bottomrule
\end{tabular}
\end{table}
The detailed simulation results are plotted in Figures \ref{fig:1}-\ref{fig:6}. Figures \ref{fig:1}-\ref{fig:2} show the results for MNB and we can see that the best results were obtained for individual character-level N-Grams extracted using Feature Count. 

Figures \ref{fig:3}-\ref{fig:4} show that the best results for SVM, for both combined and individual N-Gram, were obtained using character-level N-Grams extracted using TF-IDF. 

Figures \ref{fig:5}-\ref{fig:6} show the results obtained after using further deeper word-normalization algorithms for comparison. Deeper NLP algorithms like Porter's-Stemming and POS-Lemmatization generally require extra information about the language to normalize the word to its root form\cite{wnorm2}. 

From these results we can conclude that our research question is answered in the affirmative, and that character-level N-Grams are a superior choice for micro-text classification tasks.
\section{Conclusions}
We show that feature selection of character-level groups ranging from (1,6) to (1,8) provide the highest accuracy when extracted using TF-IDF metrics and trained on an SVM classifier. MNB classifiers generally tend to perform better when trained using FC at the character level.

We also observe that further pre-processing by Lemmatization and Stemming did not result in any significant gains, which corroborates previous work in this area \cite{wnorm}.

We conjecture that $\sim$65\% accuracy is near the maximum that can be achieved for this task, since this is a multi-label task with many features and a sparse dataset.

We hypothesize that character-level features may help in omitting various issues found with word-level features, such as spelling errors, tenses, singularization/pluralization, etc. This approach may allow us to avoid the overhead of normalizing all words by Stemming/Lemmatization. While using character-level features restricts our use only to the group of characters, it also permits us to work without knowledge of the language, which is required for NLP algorithms.

Overall, we observe that the character-level results tend to be superior. As a result, we conclude that character-level feature selection is a feasible and effective step for multi-class textual classification of tweets.

\section{Future Work}
Because character-level featurization appears to perform well, we may be able to obtain similar results without the need to clean the dataset.  We intend to test this hypothesis in future work. We also plan to test our method on binary classification problems, such as sentiment analysis, to determine the effect of character-level features on accuracy. Comparison with Deep Learning methods like Convolutional Neural Networks and Recurrent Neural Networks, which have been recently found to perform well for textual classification tasks on unstructured text corpora \cite{fw1,fw2}, is another potential area for further study.